# Intralayer Negative Poisson's Ratio in Two-Dimensional Black Arsenic by Strain Engineering

Jingjing Zhang,[1,2,#] Weihan Zhang,[3,#] Leining Zhang,[4,#] Guoshuai Du,[1,2] Yunfei Yu,[1,2] Qinglin Xia,[5] Xu Wu,[1,6] Yeliang Wang,[6] Wei Ji,[3] Jingsi Qiao,[1,6,*] Feng Ding,[7,*] Yabin Chen[1,2,*]

[1]Advanced Research Institute of Multidisciplinary Sciences, Beijing Institute of Technology, Beijing 100081, P.R. China

[2]School of Aerospace Engineering, Beijing Institute of Technology, Beijing 100081, P.R. China

[3]Beijing Key Laboratory of Optoelectronic Functional Materials & Micro-Nano Devices, Department of Physics, Renmin University of China, Beijing 100872, P.R. China

[4]Beijing Key Laboratory of Construction Tailorable Advanced Functional Materials and Green Applications, MOE Key Laboratory of Cluster Science, School of Chemistry and Chemical Engineering, Beijing Institute of Technology, Beijing 100081, P.R. China

[5]School of Physics and Electronics, Hunan Key Laboratory of Nanophotonics and Devices, Central South University, Changsha 410083, P.R. China

[6]MIIT Key Laboratory for Low-Dimensional Quantum Structure and Devices, School of Information and Electronics, Beijing Institute of Technology, Beijing 100081, P.R. China

[7]Faculty of Materials Science and Engineering/Institute of Technology for Carbon Neutrality, Shenzhen Institute of Advanced Technology, Chinese Academy of Sciences, Shenzhen 518055, P.R. China

[*]Corresponding authors: J.Q. (qiaojs@bit.edu.cn), F.D. (F.ding@siat.ac.cn), Y.C. (chyb0422@bit.edu.cn)

[#]These authors contributed equally to this work.

**ABSTRACT:** Negative Poisson's ratio as the anomalous characteristic generally exists in artificial architectures, such as re-entrant and honeycomb structures. The structures with negative Poisson's ratio have attracted intensive attention due to their unique auxetic effect and many promising applications in shear resistant and energy absorption fields. However, experimental observation of negative Poisson's ratio in natural materials barely happened, although various two-dimensional layered materials are predicted in theory. Herein, we report the anisotropic Raman response and the intrinsic intralayer negative Poisson's ratio of two-dimensional natural black arsenic (b-As) via strain engineering strategy. The results were evident by the detailed Raman spectrum of b-As under uniaxial strain together with density functional theory calculations. It is found that b-As was softer along the armchair than zigzag direction. The anisotropic mechanical features and van der Waals interactions play essential roles in strain-dependent Raman shifts and negative Poisson's ratio in the natural b-As along zigzag direction. This work may shed a light on the mechanical properties and potential applications of two-dimensional puckered materials.

**KEYWORDS:** Black arsenic; Strain engineering; Negative Poisson's ratio; Puckered structure; Raman spectroscopy; Van der Waals interaction

## INTRODUCTION

Poisson's ratio $v$, as a fundamental mechanical property, is defined as the ratio of the induced transverse strain to axial strain.[1, 2] For the isotropic materials, $v$ can be principally expressed by $v = [3(B/G-2)]/[6(B/G+2)]$, where bulk modulus $B$ and shear modulus $G$ are associate with the volumetric and transverse deformations, respectively.[3] This relation essentially defines the numerical limits for Poisson's ratio, $-1 \leq v \leq 1/2$ for $0 \leq B/G < \infty$, and most conventional materials generally have a positive Poisson's ratio. Negative Poisson's ratio (auxetic effect), an anomalous and interesting phenomenon, enables the lateral expansion instead of contraction when stretched lengthwise. With the prominent shear resistance, fracture toughness and shock absorption, this specific kind of auxetic materials possesses superior mechanical properties and are widely used in many practical fields, such as biomedical engineering, aerospace, sensor devices and novel functional structures.[4-6] The negative Poisson's ratio effect generally results from the special structure and deformation mechanism.[6-8] For decades, many efforts are extensively attracted to construct artificial structures with auxetic effect, including the re-entrant polymer foams ($v$ = −0.7),[9] honeycomb structures ($v$ = −0.5),[10, 11] and carbon nanotube networks ($v$ = −0.2).[12]

It is with enormous challenge to explore a natural material with auxetic properties. Up to now, two-dimensional (2D) layered materials have drawn considerable attention and exhibited many unique mechanical, optical and electrical properties, including their super flexibility, thickness-dependent band structures, and quantum transport.[13-20] Their excellent mechanical properties of 2D materials allow us to demonstrate the strain engineering to modulate its lattice structure, carrier mobility and electron-phonon coupling.[21-25] Importantly, it was found that many 2D materials, especially with puckered structures, can present negative Poisson's ratio, which can be categorized as two types: interlayer negative Poisson's ratio and intralayer negative Poisson's ratio.[26-31] For instance, group IV monochalcogenides (SiS, GeS, SnTe, etc.) are theoretically predicted to show cross-plane negative Poisson's ratio, this is, lattice constant $z$ expands when stretching the structure along $x$ direction.[32] Moreover, 2D elemental materials in group VA displayed the controversial Poisson's ratio properties. Raman spectrum and theoretical simulation by Y. Du *et al.* demonstrated the auxetic effect of bulk black phosphorus (b-P) with both the obvious cross-plane interlayer and intralayer negative Poisson's ratio, through applying uniaxial strain along armchair (AC) direction.[26] In contrast, single layer b-P with hinged structure even presented the cross-plane negative Poisson's ratio ($v = -0.027$) when the uniaxial deformation along zigzag (ZZ) direction.[30]

2D black arsenic (b-As), as a cousin of b-P, has emerged recently and attracted extensive attention owing to its more stability and extremely anisotropic properties.[33, 34] Its band gap ranges from ~0.3 eV (bulk) to ~1.4 eV (monolayer),[35-37] and the carrier's mobility is as high as $10^3$ $cm^2V^{-1}\cdot s^{-1}$ magnitude.[38, 39] Compared with b-P, theoretical calculations showed that b-As exhibits a negative Poisson's ratio of -0.09, i.e. three times larger than that of b-P.[32, 40, 41] Moreover, it is predicted that b-As can withstand a larger strain limit. Accordingly, the mechanical properties of 2D b-As strongly depends on its lattice orientation and layer number.[42-44] It was found that the Poisson's ratio of few layer b-As becomes more negative and eventually approaches the limit ($v = -0.12$) when the layer number goes to four.[29, 32] Obviously, the internal mechanism remained elusive despite much effort, and the Poisson's ratio of b-As needs to be verified experimentally due to the lack of convincing experimental evidences.

In this work, we systematically investigated the uniaxial strain responses of three Raman vibration modes ($A_g^1$, $B_{2g}$, and $A_g^2$) of b-As based on two-point bending method combined with density functional theory (DFT) calculations. It was eventually revealed that intralayer negative Poisson's ratio characteristic exists in the natural b-As under strain along ZZ direction. Raman slopes of out-of-plane $A_g^1$ mode and in-plane $B_{2g}$ and $A_g^2$ modes exhibit the distinct strain-dependence when the uniaxial strain applied along in-plane AC and ZZ axes, respectively. The detailed DFT results confirmed that b-As is softer along AC than ZZ direction. In strain engineering, anisotropic mechanical features bring different variations of bond lengths and angles, leading to opposite Raman shifts. Our results can open up new avenues on the mechanical properties and future applications of 2D b-As.

## RESULTS AND DISCUSSION

It is well known that the artificial concave architecture can usually display negative Poisson's ratio property when its concave angle reaches a critical degree, as depicted in Figure 1a. Accordingly, as the tensile stress is applied along the horizontal direction, the inclined bars rotate and the concave angles increase remarkably, accompanied by the obvious expansion in the perpendicular direction.[6, 45] Compared with this specific architecture, the natural b-As is predicted to feature negative Poisson's ratio property due to its in-plane anisotropic puckered structure in Figure 1b. In one atomic layer of b-As, each arsenic atom forms three $\sigma$ bonds with $sp^3$ hybridization, resulting in the in-plane covalent bonds longer than out-of-plane bond. This unique folded structure enables us to probe the extreme in-plane anisotropy, as reported in the orientation-dependent optical and electrical results.[36, 40, 46]

Two-point bending method[47-49] was used to apply the tunable and uniaxial strain along a given direction, where the b-As flake was located at the center of the flexible polyethylene terephthalate (PET, and its Young's modulus is as low as ~2.5 GPa) substrate. Figure 1c represents a schematic diagram of our home-made setup, which is capable of *in situ* Raman characterizations under bending conditions. Meanwhile, this versatile setup can simultaneously realize the tensile and compressive states by flipping the entire device over. Strain is efficiently transferred to b-As sample through the flexible PET substrate. A thin polyvinyl alcohol (PVA) film was normally coated on the top surface of b-As and PET, in order to protect b-As from degradation in air and to prevent the interfacial slippage or corrugation of the sample during loading test as well. Based on the fundamental mechanics, the applied strain ($\varepsilon$) can be expressed as $\varepsilon = \tau(\sin\theta)/2a$, where $\tau$ denotes the PET thickness, $\theta$ represents the angle between the tangent and horizontal directions at the endpoints of the flexible substrate, and $2a$ is the distance between two loading endpoints.[3, 50] The detailed derivation is given in Supplementary Figure S1.

Raman spectrum of 2D layered material, as a powerful means to probe the structural change and electron-phonon coupling, is quite sensitive to the external force field. Figure 1b displays three typical phonon vibration modes of b-As crystal, including the out-of-plane $A_g^1$ and the in-plane $B_{2g}$ and $A_g^2$. It is obvious that the dominant components of $B_{2g}$ and $A_g^2$ modes vibrate along the ZZ and AC directions, respectively. Crucially, the $A_g^1$ mode, describing the relative motions of arsenic atoms at the top and bottom, can rationally reveal the possible out-of-plane negative Poisson's ratio of the puckered b-As.

To evaluate the lattice orientation and crystal quality of b-As specimen, angle-resolved polarized Raman measurements and high-resolution transmission electron microscopy (HRTEM) were performed on various b-As flakes (Figure S2). First of all, large-area b-As flakes were mechanically exfoliation onto PET surface using scotch tape (see the experiment section for more details). The thickness of b-As flakes was examined by its optical contrast and also precisely determined by atomic force microscopy (AFM). Afterward, polarized Raman spectroscopy was attempted to distinguish the AC and ZZ directions of b-As. As far as we know, Raman intensity of $A_g$ modes of b-As is profoundly affected by the sample thickness and excitation wavelength, because of the anisotropic interference and absorption effects.[51] In contrast, Raman intensity of $B_{2g}$ mode remains unchanged and shows the repeated polarization dependence.[36, 52] As shown in Figure 1e, in the parallel configuration, when the two main lattice axes are aligned with the laser polarization, the $B_{2g}$ mode cannot be detected due to matrix cancellation. Therefore, it allows us to conveniently identify the ZZ or AC directions of b-As flake. In addition, the results were completely confirmed by HRTEM image and selected-area electron diffraction pattern. In Figure 1d, the orthorhombic structure is distinct and further evidenced by the lattice spacing of 4.5 and 3.7 Å along the AC and ZZ directions, respectively, well consistent with our theoretical results ($a$ = 4.46 Å, $b$ = 3.74 Å, and $c$= 11.03 Å).

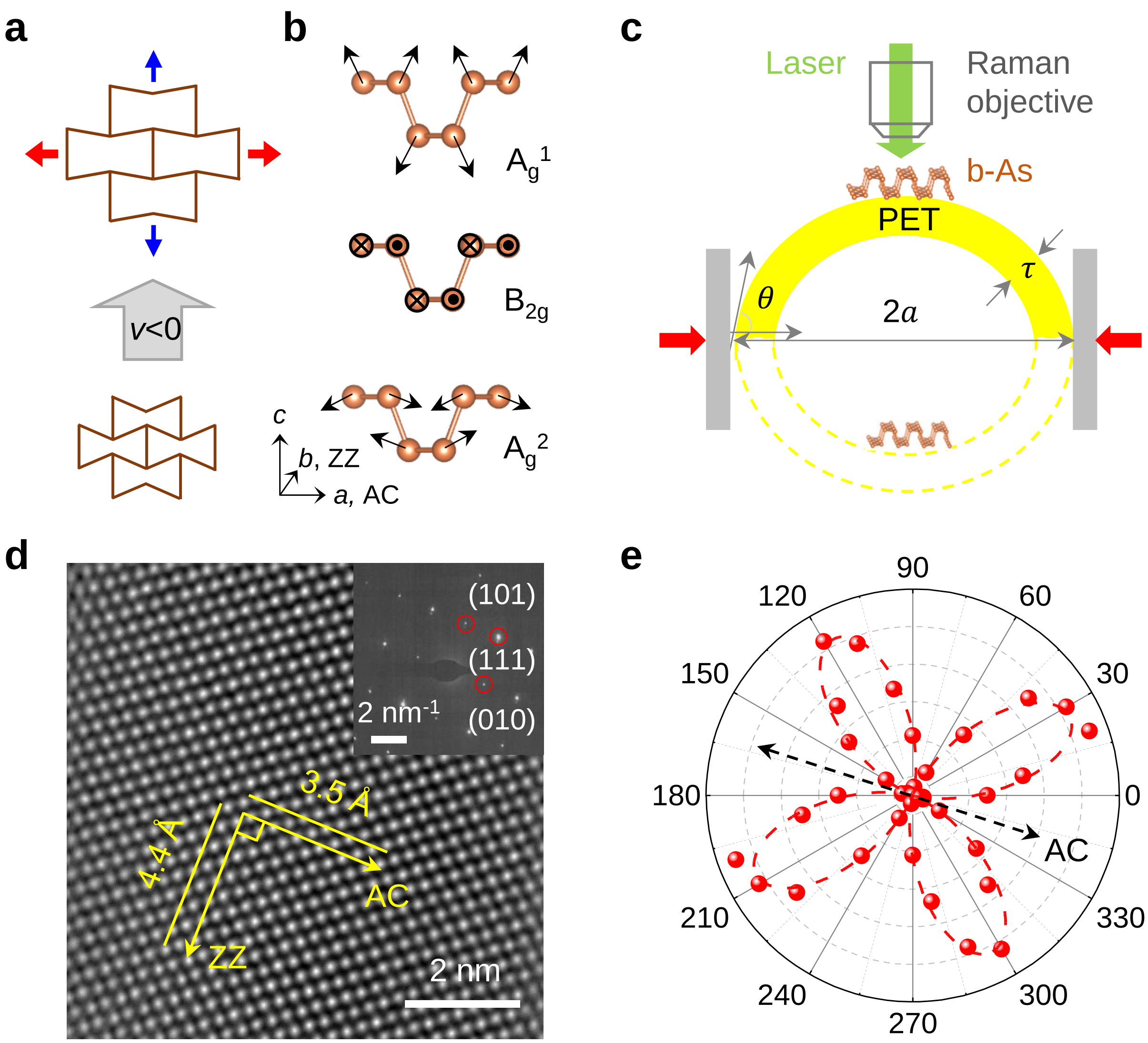

**Figure 1**. Strain engineering in b-As and its structural characterizations. (a) The typical concave architecture that basically shows a negative Poisson's ratio. The structure expands perpendicularly when the tensile strain applied along horizontal direction. (b) Schematic diagram of crystal structure of b-As and three typical Raman modes. It is noted that $A_g^1$, $B_{2g}$ and $A_g^2$ vibrate along out-of-plane, in plane ZZ and AC directions, respectively. (c) Schematic diagram of our home-made bending setup. The uniaxial strain can be tuned by bending a flexible PET substrate with a b-As flake on the top. This setup enables to apply compressive strain as well (dished lines). (d) HRTEM image of b-As with atomic resolution. The ZZ and AC directions are marked clearly. Inset is the SAED pattern. (e) Polarized Raman diagram of $B_{2g}$ mode as a function of the angle $\theta$ between laser polarization and AC direction. The red dashed line shows the fitting result with $I \propto \sin^2(2\theta)$. The black arrow indicates the AC direction.

Next, strain-dependent Raman shifts, theoretical calculations and the systematic analysis of b-As flake were demonstrated to explore the uniaxial strain-induced structural transformation and Poisson's ratio along the different AC and ZZ orientations, respectively. For the natural b-As without any deliberate strain, Raman shifts of $A_g^1$, $B_{2g}$ and $A_g^2$ modes are fitted as 219.82, 225.87, and 253.58 $cm^{-1}$ in the parallel-polarized configuration, respectively. Figure 2a shows the detailed evolution of the acquired Raman spectra of b-As under uniaxial strain along AC direction. The applied strain ranged from the tensile +0.31% to compressive -0.31%. It is noted that both $B_{2g}$ and $A_g^2$ modes show a slight blueshift as the strain gradually increases, however, $A_g^1$ mode inversely exhibits an obvious redshift. As illustrated in Figure 2b, the linear rates of $A_g^1$ mode are extracted as -2.13 and -1.86 $cm^{-1}$/% under the compressive and tensile strain, respectively. In comparison, $A_g^2$ mode hardened as increasing strain, and the corresponding rates are 0.58 (compression) and 0.50 (tension) $cm^{-1}$/%. For a specific phonon mode, the rate difference between compressive and tensile strains are primarily attributed to the non-monotonic response on the interlayer and intralayer couplings. The data points of $B_{2g}$ mode seem fairly scattered, accompanied with a negligible shift rate, which basically results from this special vibration mainly along ZZ direction, and hence it is insensitive to the external stimulus along AC direction.

When the strain was loaded along ZZ direction, Figure 2c and 2d display the significantly different results from the former. Apparently, both the in-plane $B_{2g}$ and $A_g^2$ vibration modes experience clear redshift when the tensile strain is stretched along ZZ direction, with a slope of -6.00 and -2.67 $cm^{-1}$/%, respectively. On the other hand, $B_{2g}$ and $A_g^2$ modes have a blueshift at a rate of -5.48 and -2.61 $cm^{-1}$/% as the compressive strain, as shown in Figure 2d. Although these two models exhibit the same trends, the slope of $B_{2g}$ mode is about twice larger than that of $A_g^2$ mode. It can be attributed to the fact that $A_g^2$ mode vibrates along in-plane AC direction, so it is insensitive to the strain in ZZ direction. Meanwhile, the out-of-plane $A_g^1$ mode presents a faint blueshift due to the relatively scattered data, which arises from its intrinsically weak Raman intensity in ZZ direction.

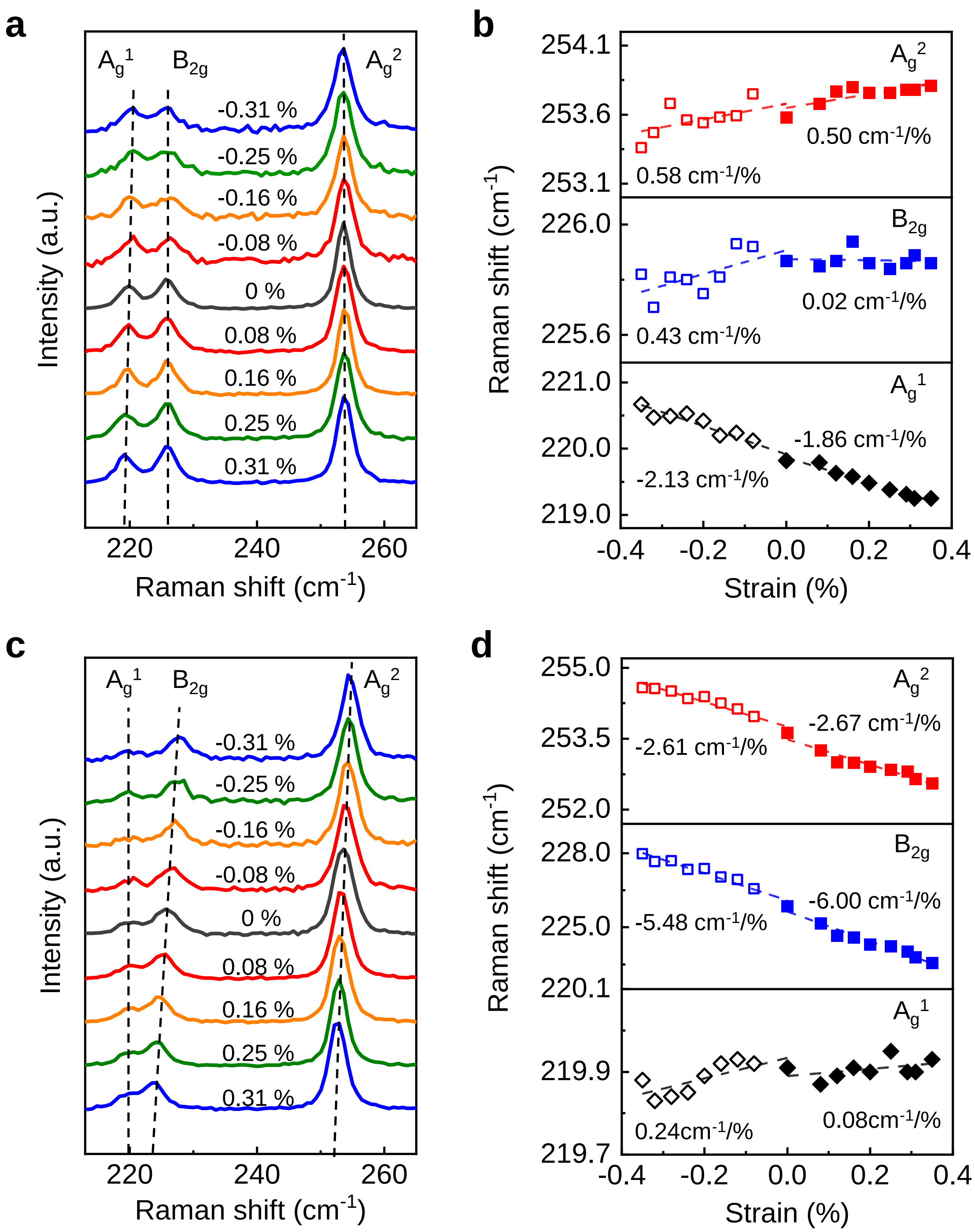

**Figure 2**. Raman spectra of b-As (15.8 nm thick) under uniaxial strain along different directions. (a) Raman spectrum evolution of b-As when the strain applied along AC direction. (b) Raman shifts of $A_g^1$, $B_{2g}$ and $A_g^2$ modes as a function of strain along AC direction. (c) Strain-dependent of Raman spectrum of b-As with strain along ZZ direction. (d) Raman shifts of three phonon modes as a function of strain along ZZ direction. Dashed lines show the linear fitting results.

To understand the Raman results and further explore its internal mechanism, we first performed DFT calculations with the optB86b to investigate the geometric and vibrational properties of strained b-As (See DFT methods for more details).[53, 54] We labeled the lattice parameters highly related to vibrational modes in Figures 3a and 3b, including three lengths ($R_1$~$R_3$), the projection distances of $R_2$ ($R_{2x}$, $R_{2z}$), and two bond angles $\theta_1$ and $\theta_2$. Based on the atomic displacements of three phonon modes in Figure 1b, their strain dependence of Raman shifts can be analyzed in depth according to the relative lattice deformation. In detail, $A_g^1$ mode reflects the cross-plane vibration of arsenic atoms as discussed above, which is mainly sensitive to intralayer bond length $R_2$ and bond angle $\theta_2$. As shown in Figures 3c and 3e, when the strain was applied along AC direction, the lattice constants of *a* and *c* shorten apparently, while bond length $R_2$ elongates, leading to the weakened interatomic interaction and the reduced Raman frequency of $A_g^1$. Instead, both in-plane $B_{2g}$ and $A_g^2$ vibrations are related to bond length $R_1$ and bond angle $\theta_1$. The blueshifts of $A_g^2$ and $B_{2g}$ modes are attributed to their enhanced As-As interactions due to the decreased in-plane bond lengths $R_1$ and bond angle $\theta_1$. Interestingly, the Raman shift trend along ZZ direction is completely opposite to that along AC direction, which results from their unique response of lattice structure to the external strain in Figure 3f-3h. In this case, bond length $R_1$ elongates and both $R_2$ and $R_3$ shorten, along with the reduced $\theta_2$ and larger $\theta_1$. In general, b-As is softer along AC than ZZ direction. Anisotropic mechanical features result in different responses of bond lengths and angles, inducing the distinct strain-dependent Raman shifts uniaxial strain engineering.

Furthermore, we comprehensively discuss the fascinating Poisson's ratio properties of b-As by DFT calculations with different functionals, including Perdew–Burke–Ernzerhof (PBE, without vdW correction),[55] optB86b-vdW and DFT-D3.[56, 57] It is well known that vdW interaction plays a vital role in physical properties of 2D layered systems, especially for Group V elemental nanomaterials.[58] After optimizing the lattice structure using optB86b-vdW functional, we found that lattice constant *c* was synergistically determined by out-of-plane projection $R_{2z}$ and interlayer distance $R_3$. The results of optB86b-vdW method showed that the interlayer distance $R_3$ remains almost constant, while both $R_{2z}$ and lattice constant *c* decrease slightly under the strain along AC direction, indicating that there is no omitted negative Poisson's ratio effect in sublayers or interlayer. These results were further verified by DFT-D3 functional in Figure S6, distinguished from PBE functional without considering vdW correction. The PBE functional displayed that both $R_3$ and lattice constant *c* increased when the strain along AC direction, suggesting an obvious interlayer negative Poisson's ratio in Figure S7. This phenomenon is controversial, although it is inclined to agree with the literature result of b-P.[26] Notably, Figure 3g illustrates that out-of-plane projection distance $R_{2z}$ increases significantly as the tensile strain applied in ZZ direction, consistent with the theoretical results of monolayer b-As,[29] so it is confirmed that the natural b-As has intralayer negative Poisson's ratio in this scenario. Importantly, this unique property should be remained as the thickness of black arsenic approaches the monolayer limit.

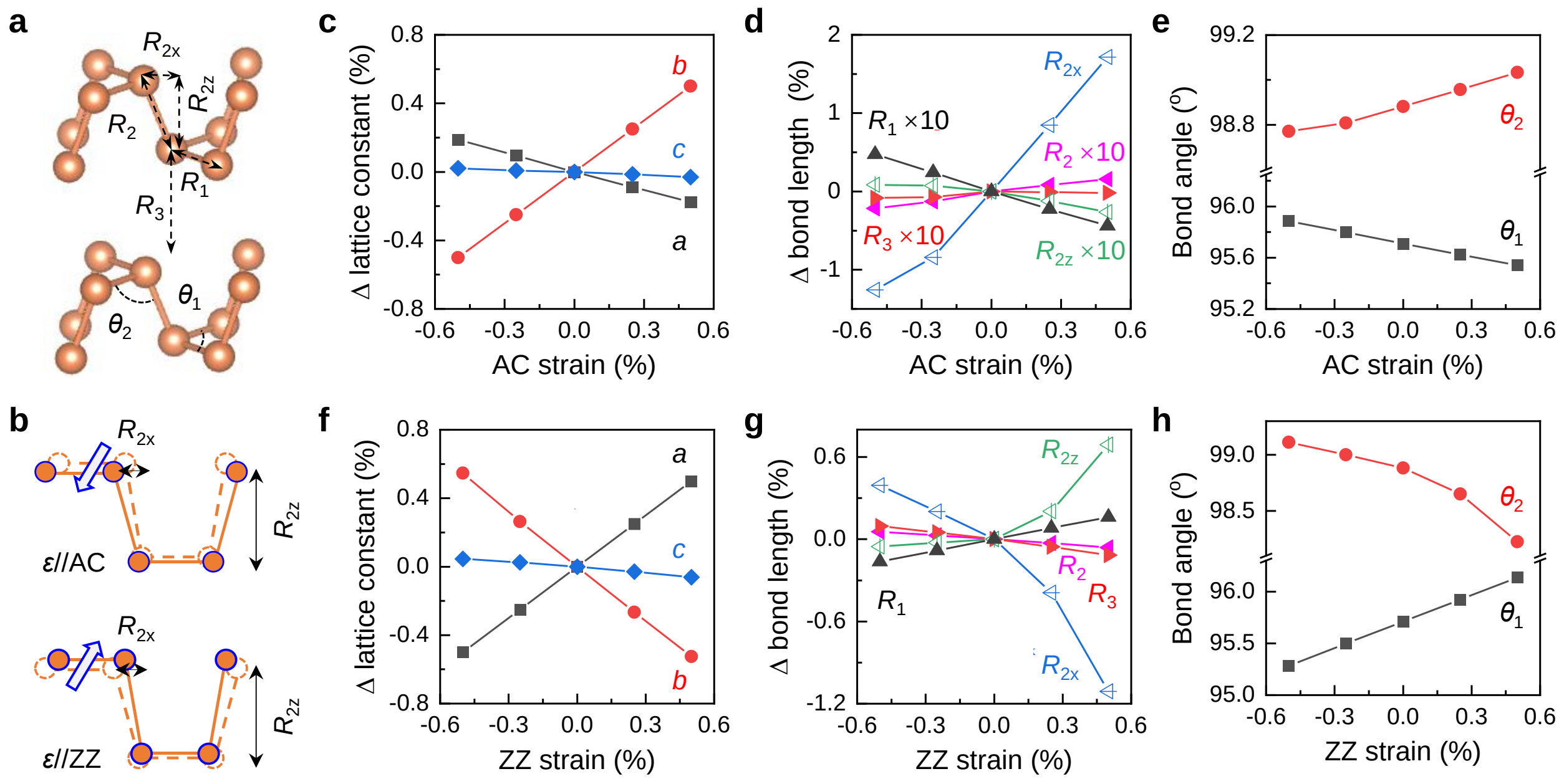

**Figure 3**. Intralayer negative Poisson's ratio and the microscopic mechanics in 2D b-As when the strain was applied along ZZ direction. (a) Atomic structure of b-As flake with the labeled bond lengths and angles. (b) Schematic illustration of lattice deformations of b-As under uniaxial strain along AC (top) and ZZ (bottom) directions. (c) Lattice constants *l* evolution of b-As as strain along AC direction. $l\% = (l_s - l_0)/l_0$, where $l_s$ and $l_0$ denote the lattice constant with and without strain for *a*, *b*, and *c*, respectively. (d) The normalized bond lengths vary as the strain along AC direction. $R\% = (R_s - R_0)/R_0$, where $R_s$ and $R_0$ represent the bond length with and without strain, respectively. (e) Strain-dependent bond angles of b-As along AC direction. (f) Lattice constants evolution of b-As as strain along ZZ direction. (g) The normalized bond lengths vary as the strain along ZZ direction. (h) Strain-dependent bond angles of b-As along ZZ direction.

Subsequently, thickness-dependent Raman shift rates in $A_g^1$, $B_{2g}$ and $A_g^2$ modes of b-As were further investigated under the same strain conditon. After the bending test, the PVA film was removed by DI water and the sample thickness was accurately measured by AFM technique, as show in Figure S2. In Figure 4, it can be seen that all $A_g^1$, $B_{2g}$ and $A_g^2$ modes clearly present the similar anisotropic trends along different directions. When the tensile or compressive strain was along ZZ direction, the Raman slopes of in-plane $B_{2g}$ and $A_g^2$ modes decrease obviously as the flake thickness increases. Raman slope $K$ is definied as $K = \partial\omega/\partial\varepsilon$, where $\omega$ and $\varepsilon$ are the Raman shift and the applied strain, respectively. It can be explained that mechanical properties of thiner b-As are less affected by the interlayer coupling. However, the results along AC direction present little difference with the flake thickness, and the fluctuation of experimental data is significant.

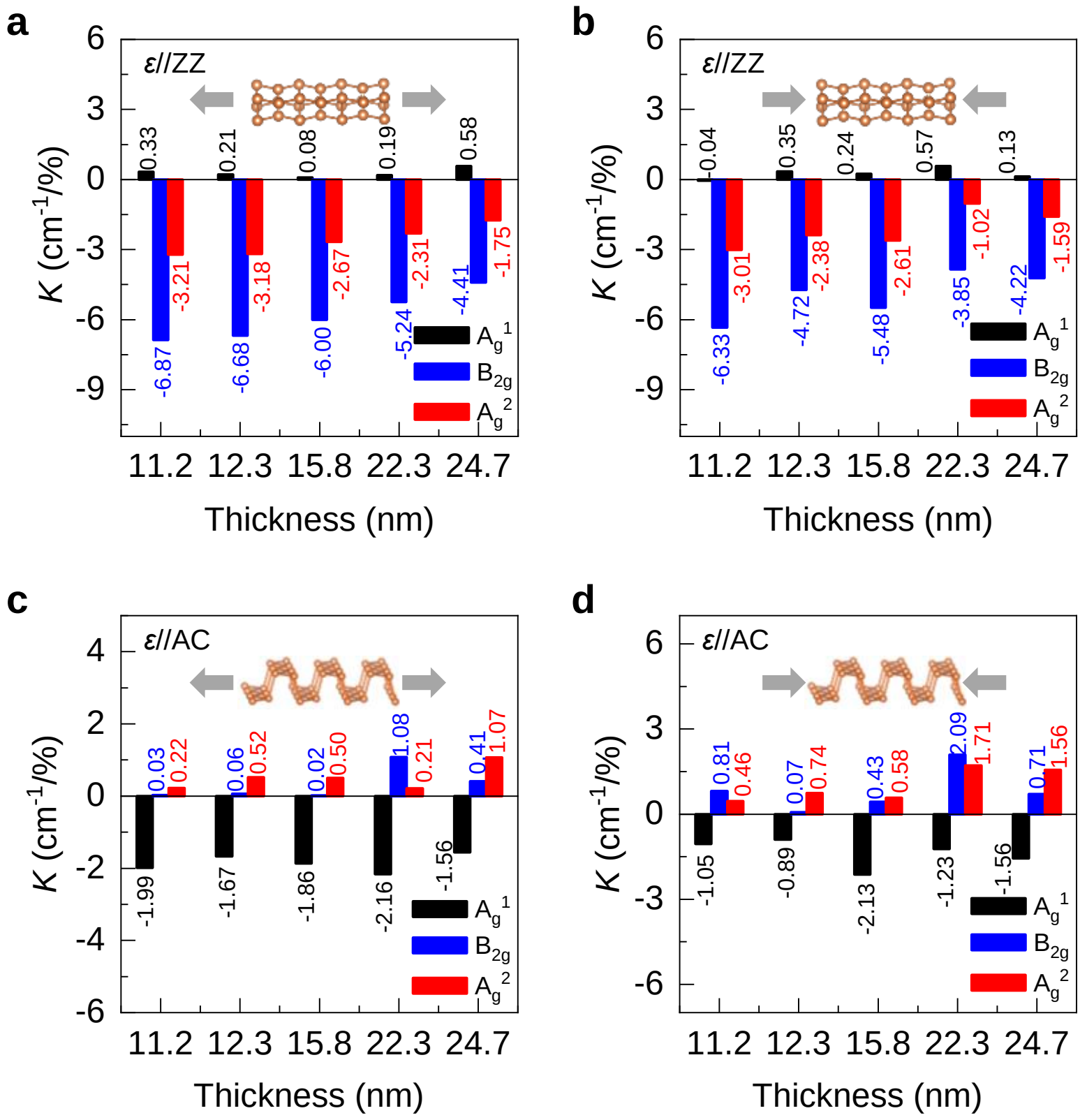

**Figure 4**. Thickness-dependent Raman responses of b-As sample under uniaxial strain. (a) Thickness dependence of Raman shift rates ($K = \partial\omega/\partial\varepsilon$) of three modes under tensile strain along ZZ direction. (b) Thickness dependence of Raman shift rates under compressive strain along ZZ direction. (c) Thickness dependence of Raman shift rates of three modes under tensile strain along AC direction. (d) Thickness dependence of Raman shift rates under compressive strain along AC direction.

## CONCLUSIONS

In summary, we have demonstrated the intralayer negative Poisson's ratio property in the natural 2D b-As with puckered structure, evidented by the unaxial-stress regulated Raman spectrum and the detailed DFT calculations. The strain engineering based on two-point bending method proved the anisotropic response of $A_g^1$, $B_{2g}$ and $A_g^2$ phonon modes when strain applied along AC and ZZ directions, well consistent with our theoretical results. Intralayer negative Poisson's ratio of b-As was confirmed with the anomalous Raman response to both compressive and tensile strain along ZZ direction. DFT calculations clarified that anisotropic mechanical features induced different responses of bond lengths and angle, and further vdW interaction played the critical role in Poisson's ratio characteristic. We believe these results could boost the research on mechanical properties of anisotropic 2D materials and futher promote their wide applications.

## EXPERIMETAL METHODS

### Preparation and Characterization of B-As Flake

The b-As sample used in this work was cleaved from the natural mineral. Then, the b-As flake with different thickness was mechanically exfoliated to a flexible PET substrate using Scotch tape. The PET syubstate was shaped to square (the dimension of 25 mm × 25 mm, and about 230 μm thick) and the b-As was located at the center of PET surface, which are convenient to the altter strain engineering process. Combined with the transmission and reflection observations of optical microscope, b-As thickness can be roughly estimated. AFM was used to accurately determine its thickness after the bending test. Importantly, a PVA coating was essential to prevent oxidation and slippage of b-As flakes, and the PVA layer can be easily washed away with DI water.

### Raman Spectrum Measurements

All Raman measurements were performed on the Horiba iHR550 system. 633 nm laser was used to excite the angle-resolved polarized Raman spectrum. For polarized Raman measurements, the analyzer was set parallel to the incident polarization direction of the laser. Then the angler-resolved polarization Raman spectrum was acquired by rotating the sample from 0-180°. Raman spectrum under uniaxial strain was measured using 532 nm laser. All characterizations were performed with a100X objective lens and a grating of 1800 grooves $mm^{-1}$. The laser beam was focused as about 1 μm and spectral resolution was better than 0.1 $cm^{-1}$. The weak laser power as ~20 μW was used to protect the b-As sample from the potential heating effect.

### DFT Calculations

DFT calculations were performed using the generalized gradient approximation for the exchange-correlation potential, the projector augmented wave method and a plane-wave basis set as implemented in the Vienna ab initio simulation package.[59-61] Density functional perturbation theory was employed to calculate vibrational frequencies at the G point.[62] The van der Waals-density functional method was used to describe the vdW interactions together with the optB86b exchange function,[54, 63] which is proved to be accurate in reproducing atomic structure of layered materials.[64] The energy cut-off for the plane-wave basis was set to 700 eV for all calculations. A *k*-mesh of 31×31×15 was adopted to sample the first Brillouin zone of the conventional unit cell of bulk b-As in geometric optimization and phonon frequencies. The shape of each cell was optimized fully and all atoms in the cell were allowed to relax until the residual force per atom was less than $1\times10^{-4}$ eV/Å. For PBE and DFT-D3 functionals, the calculations were performed with the projector augmented wave method.[61] The exchange–correlation functions were treated by the PBE generalized gradient approximation.[55] Despite the PBE functional described the chemical bonding between arsenic atoms well, it normally ignores the dispersion interaction. In addition, the DFT-D3 functional containing the long-range vdW interactions was also applied.[56, 57] To optimize the lattice constant of the unit cell, the force on each atom of 0.01 eV $Å^{-1}$ and energy convergence of $10^{-4}$ eV were set as the criteria, companying with the energy cutoff of 500 eV.

## ASSOCIATED CONTENT

### Supporting Information

The Supporting Information is available free of charge on the website.

Optical pictures of the home-made setup for strain engineering of b-As, schematic diagram of strain calculation method based on two-point bending model, structure and thickness characterizations of b-As sample, optical images demonstrate the whole compressive and tensile process, strain-regulated Raman spectrum of b-As (11.2 nm thick) under uniaxial strain, the additional theoretical calculation results with PBE, DFT-D3, and optB86b-vdW functionals.

### Author Contributions

Y.C. and J.Z. conceived this research project and designed the experiment. W.J., J.Q. and W.Z. carried out the DFT calculations (optB86b-vdW functional) and data analysis. L.Z. and F.D. carried out DFT calculations using DFT-D3 and PBE functionals. J.Z. and G.D. performed the Raman measurement and AFM characterizations. J.Z., F.Y. X.W., Y.W. and Q.X. prepared all the samples. Y.C. and J.Z. wrote the manuscript with the necessary input of all authors. All authors have given approval to the final manuscript.

**Notes**

The authors declare no competing financial interests.

**Acknowledgements**

This work was financially supported by the National Natural Science Foundation of China (grant numbers 52072032, 12274467, 61761166009, 11974422, 92163206, and 62171035), the 173-JCJQ program (grant No. 2021-JCJQ-JJ-0159), the Ministry of Science and Technology (MOST) of China (grant No. 2018YFE0202700), and the Beijing Nova Program from Beijing Municipal Science & Technology Commission (Z211100002121072). Calculations were performed at Beijing Super Cloud Computing Center and the Physics Lab of High-Performance Computing of Renmin University of China.

## REFERENCES

(1) Greaves, G. N.; Greer, A. L.; Lakes, R. S.; Rouxel, T., Poisson's ratio and modern materials. *Nat. Mater.* **2011,** *10*, 823-837.

(2) Augustus, L., *A treatise on the mathematical theory of elasticity*. Cambridge University Press: 2013.

(3) Fung, Y.C.; Tong, P.; Chen, X., *Classical and computational solid mechanics*. World Scientific Publishing Company: 2017.

(4) Friis, E. A.; Lakes, R. S.; Park, J. B., Negative Poisson's ratio polymeric and metallic foams. *J. Mater. Sci.* **1988,** *23*, 4406-4414.

(5) Choi, J. B.; Lakes, R. S., Fracture toughness of re-entrant foam materials with a negative Poisson's ratio: experiment and analysis. *Int. J. Fract.* **1996,** *80*, 73-83.

(6) Lakes, R. S., Deformation mechanisms in negative Poisson's ratio materials: structural aspects. *J. Mater. Sci.* **1991,** *26*, 2287-2292.

(7) Park, S.-I.; Ahn, J.H.; Feng, X.; Wang, S.; Huang, Y.; Rogers, J. A., Theoretical and experimental studies of bending of inorganic electronic materials on plastic substrates. *Adv. Funct. Mater.* **2008,** *18*, 2673-2684.

(8) Scarpa, F., Auxetic materials for bioprostheses [In the Spotlight]. *IEEE Signal Process. Mag.* **2008,** *25*, 128-126.

(9) Lakes, R., Foam structures with a negative Poisson's ratio. *Science* **1987,** *235*, 1038-1040.

(10) Milton, G. W., Composite materials with poisson's ratios close to –1. *J. Mech. Phys. Solids* **1992,** *40*, 1105-1137.

(11) Yeganeh-Haeri, A.; Weidner, D. J.; Parise, J. B., Elasticity of α-cristobalite: A silicon dioxide with a negative Poisson's ratio. *Science* **1992,** *257*, 650-652.

(12) Hall, L. J.; Coluci, V. R.; Galvão, D. S.; Kozlov, M. E.; Zhang, M.; Dantas, S. O.; Baughman, R. H., Sign change of Poisson's ratio for carbon nanotube sheets. *Science* **2008,** *320*, 504-507.

(13) Wang, Q. H.; Kalantar-Zadeh, K.; Kis, A.; Coleman, J. N.; Strano, M. S., Electronics and optoelectronics of two-dimensional transition metal dichalcogenides. *Nat. Nanotechnol.* **2012,** *7*, 699-712.

(14) Akinwande, D.; Huyghebaert, C.; Wang, C.-H.; Serna, M. I.; Goossens, S.; Li, L.-J.; Wong, H.-S. P.; Koppens, F. H. J. N., Graphene and two-dimensional materials for silicon technology. *Nature* **2019,** *573*, 507-518.

(15) Yoffe, A., Low-dimensional systems: quantum size effects and electronic properties of semiconductor microcrystallites (zero-dimensional systems) and some quasi-two-dimensional systems. *Adv. Phys.* **2002,** *51*, 799-890.

(16) Geim, A. K.; Grigorieva, I. V., Van der Waals heterostructures. *Nature* **2013,** *499*, 419-425.

(17) Liu, Y.; Weiss, N. O.; Duan, X.; Cheng, H.-C.; Huang, Y.; Duan, X., Van der Waals heterostructures and devices. *Nat. Rev. Mater.* **2016,** *1*, 1-17.

(18) Wang, X.; Hu, Y.; Mo, J.; Zhang, J.; Wang, Z.; Wei, W.; Li, H.; Xu, Y.; Ma, J.; Zhao, J.; Jin, Z.; Guo, Z., Arsenene: A potential therapeutic agent for acute promyelocytic leukaemia cells by acting on nuclear proteins. *Angew. Chem. Int. Ed.* **2020,** *59*, 5151-5158.

(19) Hu, Y.; Qi, Z.-H.; Lu, J.; Chen, R.; Zou, M.; Chen, T.; Zhang, W.; Wang, Y.; Xue, X.; Ma, J.; Jin, Z., Van der Waals epitaxial growth and interfacial passivation of two-dimensional single-crystalline few-layer gray arsenic nanoflakes. *Chem. Mater.* **2019,** *31*, 4524-4535.

(20) Hu, Y.; Wang, X.; Qi, Z.; Wan, S.; Liang, J.; Jia, Q.; Hong, D.; Tian, Y.; Ma, J.; Tie, Z.; Jin, Z., Wet chemistry vitrification and metal-to-semiconductor transition of 2D gray arsenene nanoflakes. *Adv. Funct. Mater.* **2021,** *31*, 2106529.

(21) Jariwala, D.; Marks, T. J.; Hersam, M. C., Mixed-dimensional van der Waals heterostructures. *Nat. Mater.* **2017,** *16*, 170-181.

(22) Glavin, N. R.; Rao, R.; Varshney, V.; Bianco, E.; Apte, A.; Roy, A.; Ringe, E.; Ajayan, P. M., Emerging applications of elemental 2D materials. *Adv. Mater.* **2020,** *32*, 1904302.

(23) Liu, Y.; Huang, Y.; Duan, X., Van der Waals integration before and beyond two-dimensional materials. *Nature* **2019,** *567*, 323-333.

(24) Qin, H.; Sun, Y.; Liu, J. Z.; Li, M.; Liu, Y., Negative Poisson's ratio in rippled graphene. *Nanoscale* **2017,** *9*, 4135-4142.

(25) Wan, J.; Jiang, J.W.; Park, H. S., Negative Poisson's ratio in graphene oxide. *Nanoscale* **2017,** *9*, 4007-4012.

(26) Du, Y.; Maassen, J.; Wu, W.; Luo, Z.; Xu, X.; Ye, P. D., Auxetic black phosphorus: A 2D material with negative Poisson's ratio. *Nano Lett.* **2016,** *16*, 6701-6708.

(27) Fei, R.; Yang, L., Lattice vibrational modes and Raman scattering spectra of strained phosphorene. *Appl. Phys. Lett.* **2014,** *105*, 083120.

(28) Gao, Y.; Wen, M.; Zhang, X.; Wu, F.; Xia, Q.; Wu, H.; Dong, H., Factors affecting the negative Poisson's ratio of black phosphorus and black arsenic: electronic effects. *Phys. Chem. Chem. Phys.* **2021,** *23*, 3441-3446.

(29) Han, J.; Xie, J.; Zhang, Z.; Yang, D.; Si, M.; Xue, D., Negative Poisson's ratios in few-layer orthorhombic arsenic: First-principles calculations. *Appl. Phys. Express* **2015,** *8*, 041801.

(30) Jiang, J.-W.; Park, H. S., Negative poisson's ratio in single-layer black phosphorus. *Nat. Commun.* **2014,** *5*, 4727.

(31) Yu, L.; Yan, Q.; Ruzsinszky, A., Negative Poisson's ratio in 1T-type crystalline two-dimensional transition metal dichalcogenides. *Nat. Commun.* **2017,** *8*, 15224.

(32) Liu, B.; Niu, M.; Fu, J.; Xi, Z.; Lei, M.; Quhe, R., Negative Poisson's ratio in puckered two-dimensional materials. *Phys. Rev. Mater.* **2019,** *3*, 054002.

(33) Hu, Y.; Liang, J.; Xia, Y.; Zhao, C.; Jiang, M.; Ma, J.; Tie, Z.; Jin, Z., 2D arsenene and arsenic materials: Fundamental properties, preparation, and applications. *Small* **2022,** *18*, 2104556.

(34) Liang, J.; Hu, Y.; Zhang, K.; Wang, Y.; Song, X.; Tao, A.; Liu, Y.; Jin, Z., 2D layered black arsenic-phosphorus materials: Synthesis, properties, and device applications. *Nano Res.* **2022,** *15*, 3737-3752.

(35) Kamal, C.; Ezawa, M., Arsenene: Two-dimensional buckled and puckered honeycomb arsenic systems. *Phys. Rev. B* **2015,** *91*, 085423.

(36) Chen, Y.; Chen, C.; Kealhofer, R.; Liu, H.; Yuan, Z.; Jiang, L.; Suh, J.; Park, J.; Ko, C.; Choe, H. S.; Avila, J.; Zhong, M.; Wei, Z.; Li, J.; Li, S.; Gao, H.; Liu, Y.; Analytis, J.; Xia, Q.; Asensio, M. C., et al., Black arsenic: A layered semiconductor with extreme in-plane anisotropy. *Adv. Mater.* **2018,** *30*, 1800754.

(37) Zhang, Z.; Xie, J.; Yang, D.; Wang, Y.; Si, M.; Xue, D., Manifestation of unexpected semiconducting properties in few-layer orthorhombic arsenene. *Appl. Phys. Express* **2015,** *8*, 055201.

(38) Zhong, M.; Xia, Q.; Pan, L.; Liu, Y.; Chen, Y.; Deng, H.-X.; Li, J.; Wei, Z., Thickness-dependent carrier transport characteristics of a new 2D elemental semiconductor: black arsenic. *Adv. Funct. Mater.* **2018,** *28*, 1802581.

(39) Sheng, F.; Hua, C.; Cheng, M.; Hu, J.; Sun, X.; Tao, Q.; Lu, H.; Lu, Y.; Zhong, M.; Watanabe, K.; Taniguchi, T.; Xia, Q.; Xu, Z.A.; Zheng, Y., Rashba valleys and quantum Hall states in few-layer black arsenic. *Nature* **2021,** *593*, 56-60.

(40) Kou, L.; Ma, Y.; Tan, X.; Frauenheim, T.; Du, A.; Smith, S., Structural and electronic properties of layered arsenic and antimony arsenide. *J. Phys. Chem. C* **2015,** *119*, 6918-6922.

(41) Wang, C.; Xia, Q.; Nie, Y.; Rahman, M.; Guo, G., Strain engineering band gap, effective mass and anisotropic Dirac-like cone in monolayer arsenene. *AIP Adv.* **2016,** *6*, 035204.

(42) Lee, C.; Wei, X.; Kysar, J. W.; Hone, J., Measurement of the elastic properties and intrinsic strength of monolayer graphene. *Science* **2008,** *321*, 385-388.

(43) Sun, Y.; Liu, K., Strain engineering in functional 2-dimensional materials. *J. Appl. Phys.* **2019,** *125*, 082402.

(44) Akinwande, D.; Brennan, C. J.; Bunch, J. S.; Egberts, P.; Felts, J. R.; Gao, H.; Huang, R.; Kim, J.-S.; Li, T.; Li, Y.; Liechti, K. M.; Lu, N.; Park, H. S.; Reed, E. J.; Wang, P.; Yakobson, B. I.; Zhang, T.; Zhang, Y.W.; Zhou, Y.; Zhu, Y., A review on mechanics and mechanical properties of 2D materials-Graphene and beyond. *Extreme Mech. Lett.* **2017,** *13*, 42-77.

(45) Grima, J. N.; Winczewski, S.; Mizzi, L.; Grech, M. C.; Cauchi, R.; Gatt, R.; Attard, D.; Wojciechowski, K. W.; Rybicki, J., Tailoring graphene to achieve negative Poisson's ratio properties. *Adv. Mater.* **2015,** *27*, 1455-1459.

(46) Wu, J.; Mao, N.; Xie, L.; Xu, H.; Zhang, J., Identifying the crystalline orientation of black phosphorus using angle-resolved polarized Raman spectroscopy. *Angew. Chem. Int. Ed.* **2015,** *54*, 2366-2369.

(47) Desai, S. B.; Seol, G.; Kang, J. S.; Fang, H.; Battaglia, C.; Kapadia, R.; Ager, J. W.; Guo, J.; Javey, A., Strain-induced indirect to direct bandgap transition in multilayer $WSe_2$. *Nano Lett.* **2014,** *14*, 4592-4597.

(48) Li, Y.; Hu, Z.; Lin, S.; Lai, S. K.; Ji, W.; Lau, S. P., Giant anisotropic Raman response of encapsulated ultrathin black phosphorus by uniaxial strain. *Adv. Funct. Mater.* **2017,** *27*, 1600986.

(49) Conley, H. J.; Wang, B.; Ziegler, J. I.; Haglund, R. F., Jr.; Pantelides, S. T.; Bolotin, K. I., Bandgap Engineering of Strained Monolayer and Bilayer $MoS_2$. *Nano Lett.* **2013,** *13*, 3626-3630.

(50) Zhang, R.; Cheung, R., Mechanical properties and applications of two-dimensional materials. *Two-dimensional Materials-Synthesis, Characterization Potential Applications. Rijeka, Crotia: InTech* **2016**, 219-246.

(51) Kim, J.; Lee, J.-U.; Lee, J.; Park, H. J.; Lee, Z.; Lee, C.; Cheong, H., Anomalous polarization dependence of Raman scattering and crystallographic orientation of black phosphorus. *Nanoscale* **2015,** *7*, 18708-18715.

(52) Wang, Y.; Chen, F.; Guo, X.; Liu, J.; Jiang, J.; Zheng, X.; Wang, Z.; Al-Makeen, M. M.; Ouyang, F.; Xia, Q.; Huang, H., In-plane phonon anisotropy and anharmonicity in exfoliated natural black arsenic. *J. Phys. Chem. Lett.* **2021,** *12*, 10753-10760.

(53) Dion, M.; Rydberg, H.; Schröder, E.; Langreth, D. C.; Lundqvist, B. I., Van der Waals density functional for general geometries. *Phys. Rev. Lett.* **2004,** *92*, 246401.

(54) Lee, K.; Murray, É. D.; Kong, L.; Lundqvist, B. I.; Langreth, D. C., Higher-accuracy van der Waals density functional. *Phys. Rev. B* **2010,** *82*, 081101.

(55) Perdew, J. P.; Burke, K.; Ernzerhof, M., Generalized gradient approximation made simple. *Phys. Rev. Lett.* **1996,** *77*, 3865-3868.

(56) Grimme, S.; Antony, J.; Ehrlich, S.; Krieg, H., A consistent and accurate *ab initio* parametrization of density functional dispersion correction (DFT-D) for the 94 elements H-Pu. *J. Chem. Phys.* **2010,** *132*, 154104.

(57) Grimme, S.; Ehrlich, S.; Goerigk, L., Effect of the damping function in dispersion corrected density functional theory. *J. Comput. Chem.* **2011,** *32*, 1456-1465.

(58) Li, L.; Kim, J.; Jin, C.; Ye, G. J.; Qiu, D. Y.; Da Jornada, F. H.; Shi, Z.; Chen, L.; Zhang, Z.; Yang, F.; Watanabe, K.; Taniguchi, T.; Ren, W.; Louie, S. G.; Chen, X. H.; Zhang, Y.; Wang, F., Direct observation of the layer-dependent electronic structure in phosphorene. *Nat. Nanotechnol.* **2017,** *12*, 21-25.

(59) Blöchl, P. E., Projector augmented-wave method. *Phys. Rev. B* **1994,** *50*, 17953-17979.

(60) Kresse, G.; Furthmüller, J., Efficient iterative schemes for *ab initio* total-energy calculations using a plane-wave basis set. *Phys. Rev. B* **1996,** *54*, 11169-11186.

(61) Kresse, G.; Joubert, D., From ultrasoft pseudopotentials to the projector augmented-wave method. *Phys. Rev. B* **1999,** *59*, 1758-1775.

(62) Baroni, S.; de Gironcoli, S.; Dal Corso, A.; Giannozzi, P., Phonons and related crystal properties from density-functional perturbation theory. *Rev. Mod. Phys.* **2001,** *73*, 515-562.

(63) Klimeš, J.; Bowler, D. R.; Michaelides, A., Chemical accuracy for the van der Waals density functional. *J. Phys.: Condens. Matter* **2009,** *22*, 022201.

(64) Wu, J.-B.; Hu, Z.X.; Zhang, X.; Han, W.P.; Lu, Y.; Shi, W.; Qiao, X.F.; Ijiäs, M.; Milana, S.; Ji, W.; Ferrari, A. C.; Tan, P., Interface coupling in twisted multilayer graphene by resonant Raman spectroscopy of layer breathing modes. *ACS Nano* **2015,** *9*, 7440-7449.

**For Table of Contents Only**

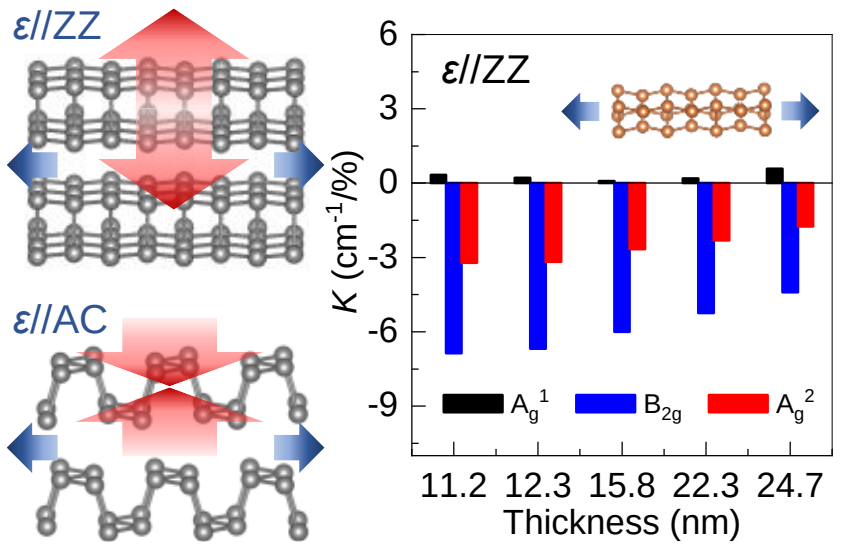

Two-dimensional black arsenic is here reported to be with the intrinsic intralayer negative Poisson's ratio via strain engineering approach. When the uniaxial strain is applied along zigzag direction, Raman measurements confirm that the red shifts of both $B_{2g}$ and $A_g^2$ modes are very different from blue shift of $A_g^1$ mode, which is well consistent with density functional theory calculations.